\pdfoutput=1

\documentclass[final,3p,times,twocolumn]{elsarticle}

\usepackage{url}
\usepackage{hyperref}

\usepackage{graphicx}
\usepackage{amssymb}

\usepackage{lineno}

\usepackage[english]{babel}

\usepackage{graphicx} 
\usepackage{color}

 \usepackage[table,xcdraw]{xcolor} 
 
\usepackage{amsmath}
\usepackage{ulem}

\newcommand{\blista}{\renewcommand{\labelenumi}{(\roman{enumi})} 
\begin{enumerate}}
\newcommand{\elista}{\end{enumerate} \renewcommand{\labelenumi}{\arabic{enumi}.}}

\newif\ifFINALVersion    
\newif\ifDIFFVersion

\FINALVersiontrue  

\ifFINALVersion
    
    \newcommand{\NEW}[1]{{#1}}
    \newcommand{\remove}[1]{{}}
\fi

\ifDIFFVersion

\newcommand{\NEW}[1]{\textcolor{blue}{#1}}
\newcommand{\remove}[1]{{}} 
\fi

\journal{Journal Name}

\begin{document}

\begin{frontmatter}


\title{Data-Driven Methods for Present and Future Pandemics:\\ Monitoring, Modelling and Managing \tnoteref{t1}}

\tnotetext[t1]{The authors belong to the CONtrol COvid-19 Team, including more than 35 researches from universities of Spain, Italy, France, Germany, United Kingdom and Argentina. The main goal of the CONCO-Team is to develop data-driven methods to better understand and control the Covid-19 pandemic. \\
\NEW{This work was supported  by the Agencia Estatal de Investigación (AEI)-Spain  under Grant PID2019-106212RB-C41/AEI/10.13039/501100011033. VP acknowledges the support of the US National Science Foundation under grants CAREER-ECCS-1651433 and NSF-III-200884556. GG acknowledges the support of the Strategic Grant MOSES at the University of Trento.}}



\author[1]{Teodoro Alamo}
\author[2]{Daniel G. Reina}
\author[3]{Pablo Millán Gata}
\author[4]{Victor M. Preciado}
\author[5]{Giulia Giordano}

\address[1]{Departamento de Ingenier\'ia de Sistemas y Autom\'atica, Universidad de Sevilla, Escuela Superior de Ingenieros, Sevilla}
\address[2]{Departamento de Ingenier\'ia Electrónica, Universidad de Sevilla, Escuela Superior de Ingenieros, Sevilla}
\address[3]{Departamento de Ingeniería, Universidad Loyola Andalucía, Seville, Spain}
\address[4]{Department of Electrical and Systems Engineering, University of Pennsylvania, Philadelphia, USA}
\address[5]{Department of Industrial Engineering, University of Trento, Trento, Italy}

\begin{abstract}
This survey analyses the role of data-driven methodologies for pandemic modelling and control. We provide a roadmap from the access to epidemiological data sources to the control of epidemic phenomena. We review the available methodologies and discuss the challenges in the development of data-driven strategies to combat the spreading of infectious diseases. Our aim is to bring together several different disciplines required to provide a holistic approach to \NEW{epidemic analysis}, such as data science, epidemiology, \NEW{and}  systems-and-control theory. A 3M-analysis is presented, whose three pillars are: Monitoring, Modelling and Managing. The focus is on the potential of data-driven schemes to address \NEW{three} different challenges raised by a pandemic: (i) monitoring the epidemic evolution and assessing the effectiveness of the adopted countermeasures; (ii) modelling and forecasting the spread of the epidemic; (iii) making timely decisions to manage, mitigate and suppress the contagion. For each step of this roadmap, we review consolidated theoretical approaches (including data-driven methodologies that have been shown to be successful in other contexts) and discuss their application to past or present epidemics, \NEW{such as Covid-19}, as well as their potential application to future epidemics.
\end{abstract}


\end{frontmatter}

\bibliographystyle{elsarticle-harv}

\section{Introduction}

The 2019 coronavirus pandemic (Covid-19) is one of the most critical public health emergencies in recent human history. While facing this pandemic, governments, public institutions, healthcare professionals, and researches of different disciplines address the problem of effectively controlling the spread of the virus while minimizing the negative effects on both the economy and society. The challenges raised by this pandemic require a holistic approach. In this document, we analyze the interplay between data science, epidemiology and control theory, which is crucial to understand and manage the spread of diseases both in human and animal populations. In line with current epidemiological needs, this paper aims to review available methodologies, while anticipating the difficulties and challenges encountered in the development of data-driven strategies to combat pandemics. We consider the Covid-19 pandemic as a case study and summarise some lessons learned from this pandemic with the hope of improving our preparedness at handling future outbreaks.

\begin{figure}[ht!]
    \centering
    \includegraphics[width=\columnwidth]{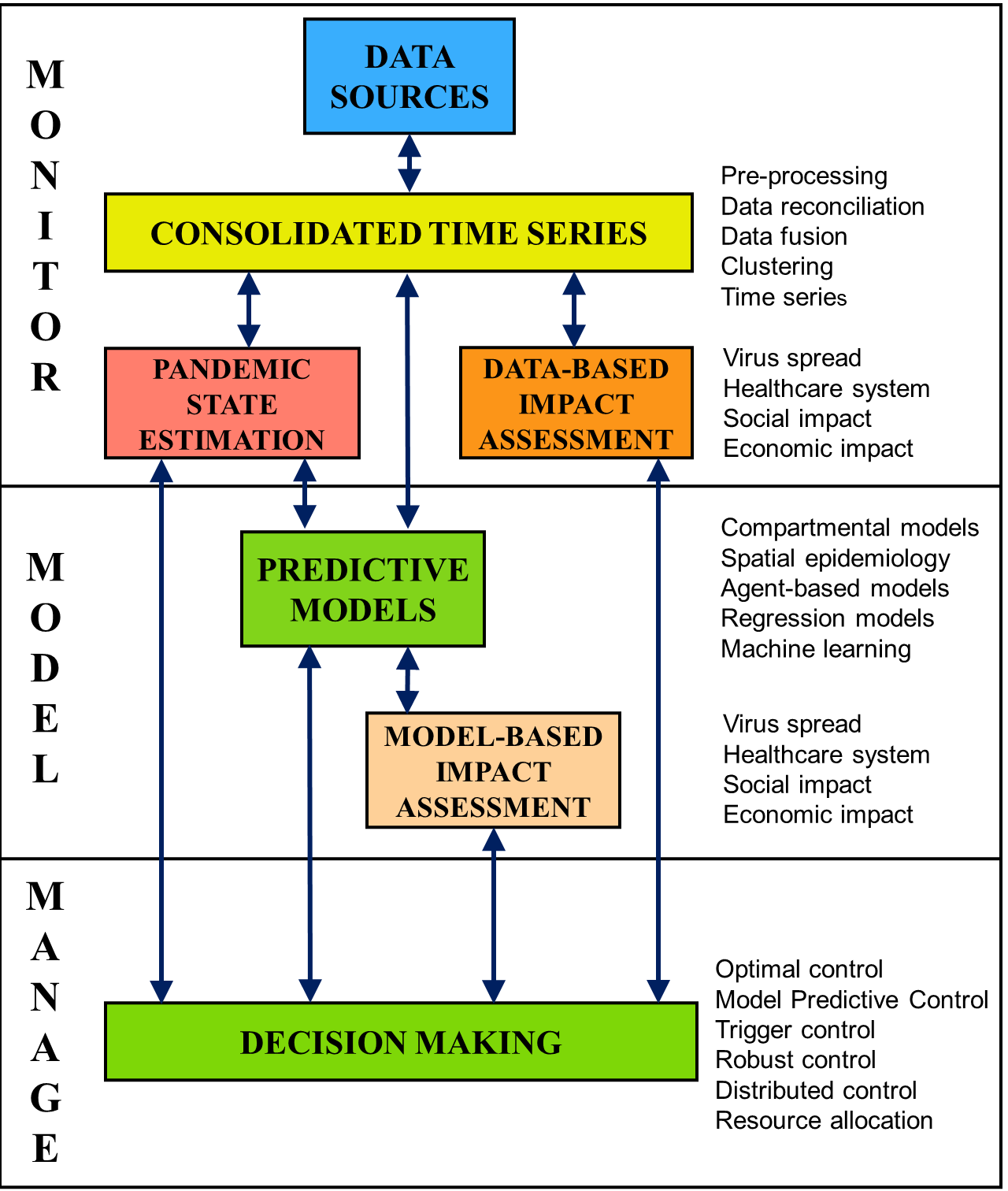}
    \caption{3M-Approach to data-driven control of an epidemic: Monitoring, Modelling and Managing.}
    \label{fig:structure}
\end{figure}

In the context of epidemics outbreaks, data-driven tools are fundamental to: (i) monitor the spread of the epidemic and assess the potential impact of adopted countermeasures, not only from a healthcare perspective but also from a socioeconomic one; (ii) model and forecast the epidemic evolution; (iii) manage the epidemic by making timely decisions to mitigate and suppress the contagion. Optimal decision making in the context of a pandemic is a complex process involving a significant amount of uncertainty; at the same time, not reacting timely and with adequate intensity, even in the presence of overwhelming uncertainties, can lead to severe consequences. This survey provides a holistic roadmap that \NEW{encompasses} from the process of retrieving epidemiological data to the decision-making process aimed at controlling, mitigating and preventing the epidemic spread. A 3M-analysis is proposed, covering three main aspects: Monitoring, Modelling and Managing, as shown in Figure \ref{fig:structure}. A more detailed document, focused on the Covid-19 pandemic, can be found in the preprint \cite{alamo2020data}.
Each step of this roadmap is presented through a review of consolidated theoretical methods and a discussion of their potential to help us understand and control pandemics.
When possible, examples of applications of these methodologies on past or current epidemics are provided. Data-driven methodologies that have proven successful in other biological contexts, or have been identified as promising solutions in the Covid-19 pandemic, are highlighted. This survey does not provide an exhaustive enumeration of methodologies, algorithms and applications. Instead, it is conceived to serve as a bridge between those disciplines required to develop a holistic approach to the epidemic, namely: data science, epidemiology, and control theory. 

Data are a fundamental pillar to understand, model, forecast, and manage many of the aspects required to provide a comprehensive response against \NEW{an epidemic, or pandemic, outbreak}. There \NEW{exist} many different open data resources and institutions providing relevant information not only in terms of specific epidemiological variables but also of other auxiliary variables that facilitate the assessment of the effectiveness of the implemented interventions (see \cite{Alamo2020b} for a review on open data resources and repositories for the Covid-19 case). Since the available epidemiological data suffer from severe limitations, methodologies to detect anomalies in the raw data and generate time-series with enhanced quality (like data reconciliation, data-fusion, data-clustering, signal processing, to name just a few) play a crucial role.

Another important aspect of the 3M-approach is the real-time surveillance of the epidemic, which can be implemented by monitoring mobility, using social media to assess the compliance to restrictions and recommendations, pro-active testing, contact-tracing, etc. The design and implementation of surveillance systems capable of early detecting secondary epidemic waves is also very important.

Modelling techniques are also fundamental in the fight against pandemics. Epidemiological models range from coarse compartmental models to complex networked and agent-based models. Fundamental parameters characterizing the dynamics of the virus can be identified using these models. Besides, data-driven parameter estimation provides mechanisms to forecast the epidemic evolution, as well as to anticipate the effectiveness of adopted interventions. However, fitting the models to the available data requires specific techniques because of critical issues like partial observation, non-linearities and non-identifiability. Sensitivity analysis, model selection and validation methodologies have to be implemented \cite{Martcheva2015}, \cite{BurnhamKennethPandAnderson2010}.  
Apart from the forecasting possibilities that epidemiological models offer, alternative forecasting techniques from the field of data science can be applied in this context. The choice ranges from simple linear parametric methods to complex deep-learning approaches. The methods can be parametric or non-parametric in nature. Some of these techniques provide probabilistic characterizations of the provided forecasts.

Several measures to mitigate the epidemic can be found in the literature, but one needs to be careful about their effectiveness \cite{Xiao2020:CONTROL:MEASURES}. Some measures, like an aggressive lockdown of an entire country, have a devastating effect on the economy and they might be adopted at very precise moments, preferably as early as possible and for short time periods. Other measures, like pro-active testing and contact-tracing, can be very effective while having a minor impact on the economy \cite{Ferretti2020}. In this direction, control theory provides a consolidated framework to formulate and solve many relevant decision-making problems \cite{Nowzari2016:ieee:CONTROL}, such as the optimal allocation of resources (e.g. test reagents and vaccines) and the determination of the optimal time to implement certain interventions. The use of optimal control theory and (distributed) model predictive control has great potential in epidemic control. Mathematical tools from the fields of control theory and dynamic systems, such as bifurcation theory and Lyapunov theory, have been extensively used to characterize the different possible qualitative behaviours of epidemics.

This survey is organized as follows:
Section \ref{sec:Estimation:State:Epidemic} describes different methodologies to monitor the current state of a pandemic. An overview of different techniques to model an epidemic is provided in Section \ref{sec:models}. The main forecasting techniques are described in Section \ref{sec:forecasting}. 
The question of how to assess the effectiveness of different non-pharmaceutical measures is analyzed in Section \ref{sec:Impact:Tools}. 
The decision making process and its link with control theory is addressed in Section \ref{sec:Decision:Making}. The review paper is finished with a section describing some conclusions and lessons learn\NEW{ed}.

\section{\NEW{Monitor--} Estimation of the state of a pandemic}\label{sec:Estimation:State:Epidemic}

There is a plethora of indicators that can be monitored in order to contain a pandemic. This includes not only estimations of the current incidence of the disease in the population and the healthcare system, but also the (daily) surveillance of measures that directly or indirectly affect its spread, such as physical distancing and mobility, as well as testing and contact tracing. 
In order to design an effective response to an epidemic outbreak, it is of \NEW{utmost} importance to build up-to-date estimations of the epidemic state. This estimation process is hindered by the presence of an incubation period of the infectious disease, which introduces a time-delay between the beginning of a new infection and its potential detection. Another challenge in the estimation process is the presence of infectious but asymptomatic cases, which is an important transmission vector in the case of \NEW{several pathogens, including HIV, Zika virus and SARS-CoV-2} \cite{Ferretti2020}. These (and other) challenges motivate the need for specific surveillance and estimation methodologies capable of using available information in order to design quick and effective control measures \NEW{\cite{Alamo:Challenges:2021}}.

In this section, we cover the most relevant techniques to monitor the state of the pandemic, focusing on approaches oriented towards (i) real-time monitoring of different aspects of the pandemic (real-time epidemiology); (ii) early detection of infected cases and immune response estimation  (pro-active testing);  (iii) estimation of the current fraction of infected population, both symptomatic and asymptomatic (state estimation methods); (iv) early detection of new waves (epidemic wave surveillance).

\subsection{Real-time epidemiology} 
The use of a large number of real-time data streams to infer the status and dynamics of a population's health presents enormous opportunities as well as significant scientific and technological challenges \cite{Bettencourt2007}, \cite{Zeng2010:BioSurveillance}, \cite{Drew2020:science:real:time:epidemiology}. 
Real-time epidemic data can vary widely in nature and origin (e.g., mobile phone data, social media data, IoT data and public health systems) \cite{Alamo2020b}, \cite{ting2020digital:data}. During the Covid-19 pandemic, mobile phone data, when used properly and carefully, have provided invaluable information for supporting public health actions across early, middle, and late-stage pandemic phases \cite{Oliver2020:MOBILE:DATA:SCIENCE}. 
Voluntary installation of Covid-19 apps or web-based tools have allowed the active retrieval of data related to exposure and infections. The information \NEW{stemming} from these sources \NEW{has} provided real-time epidemiological data that have then been used to identify hot spots for outbreaks \cite{Drew2020:science:real:time:epidemiology}. Social media have also been relevant to assess the mobility of the population and its awareness with regard to physical distancing, as well as the state of the economy and many other key indicators \cite{Zhou2020GIS:BIG:DATA}. 

Our ability to extract information regarding population mobility is essential to predict spatial transmission, identify risk areas, and decide control measures against the disease. Nowadays, the most effective tool to access real-time mobility data is through Big Data technologies and Geographic Information Systems (GIS). These systems have played a relevant role when addressing past epidemics like SARS and MERS \cite{Peeri2020:SARS:MERS:IoT}, providing efficient aggregation of multi-source big data, rapid visualization of epidemic information, spatial tracking of confirmed cases, surveillance of regional transmission and spatial segmentation of the epidemic risk \cite{Zhou2020GIS:BIG:DATA}, \cite{Wang2020BIGDATA}. 

\subsection{Proactive testing}
Proactive testing is key in the control of infectious diseases, since it allows us to identify and isolate infected individuals. It also provides relevant information to identify risk areas, fraction of asymptomatic carriers, and attained levels of immunization in the population \cite{winter2020important:SEROLOGY}, \cite{Yilmaz2020:FOCUS}. There are different methodologies to approach proactive testing:
\begin{itemize}
    \item \textbf{Risk-based approach}: In this approach, one must test first those individuals with the highest probability of being carriers of the disease (i.e. not only those with symptoms, but also those who have been heavily exposed to the disease). For example, healthcare workers are at high risk and can also be relevant transmission vectors. Second, test those individuals that have been exposed to a confirmed case according to contact tracing. Finally, test those individuals who have recently travelled to hot spots \cite{Wang2020BIGDATA}. The determination of hot spots can be done by means of government mobility surveillance or by personal software environments \cite{Drew2020:science:real:time:epidemiology}.
   \item \textbf{Voucher-based system}: In this system, people who test positive are given an anonymous voucher that they can share with a limited number of people whom they think might have infected. The recipients can use this voucher to book a test and receive their test results without ever revealing their identity. People receiving positive result are given vouchers to further backtrack the path of infection; see \cite{roomp:Oliver:2020acdc} and \cite{nanni2020:OLIVER:give} for the Covid-19 case. 
   \item \textbf{Serology studies}: Some tests (such as RT-PCR revealing viral load) are unable to detect past infection. Conversely, serological tests, carried out within the correct time frame after disease onset, can detect both active and past infections, since they detect antibodies produced in response to the disease. Serological analysis can be useful to identify clusters of cases, to retrospectively delineate transmission chains, to ascertain how long transmission has been ongoing, or to estimate the fraction of asymptomatic individuals in the population \cite{winter2020important:SEROLOGY}.
   \end{itemize}


\subsection{State-space estimation methods} 

As we will see in the next section, dynamic state-space epidemiological models are fundamental to characterize how the virus spreads in a specific region and estimate time-varying epidemiological variables that are not directly measurable \cite{Cazelles1997}, \cite{SCHARBARG:ARC:Identifiability}. Classical state-space estimation methods, like the Kalman filter \cite{Riad2019}, are employed to estimate the fraction of currently infected population. The objective of the Kalman filter is to update our knowledge about the state of the system whenever a new observation is available \cite{Durbin2012}. Different modifications and generalizations of the Kalman filter have been developed and tailored to epidemic models. These methodologies are essential both to the estimation problem and to the inference of the parameters that describe the model (see \cite{Schon2011} and \cite{Abreu2020}). 

\subsection{Epidemic wave surveillance}

Infectious diseases often lead to recurring epidemic waves interspersed with periods of low-level transmission, as observed, for example, in the \NEW{``}Spanish" flu \cite{Reid2001}, Influenza \cite{Vega2015} and Covid-19 \cite{Glass2020}. In this context, it is crucial to implement a surveillance system able to detect, or even predict, recurring epidemic waves, so as to enable an immediate response aiming to \NEW{reduce} the potential burden of the outbreak. Detecting outbreaks requires methodologies able to process huge amount of data stemming from various surveillance systems \cite{Althouse2015:SURVEILLANCE:NOVEL:DATA:STREAMS}, \cite{Dubrawski2011}, \cite{Elliot2020} and determine whether the spread of the virus has surpassed a threshold requiring mitigation measures; see, e.g. \cite{Lazarus2010:Surveillance}. A large body of literature focuses on epidemiological detection problems, since many infectious diseases undergo considerable seasonal fluctuations with peaks seriously impacting the healthcare systems \cite{Sparks2013}, \cite{Unkel2012:REVIEW:sURVEILLACE}. National surveillance systems are implemented world-wide to rapidly detect outbreaks of influenza-like illnesses, and assess the effectiveness of influenza vaccines \cite{Vega2015}, \cite{Thompson2006}. Specific methodologies to determine the baseline influenza activity and epidemic thresholds have been proposed and implemented \cite{Vega2013:MOVING:EPIDEMIC:METHOD}. These methods aim at reducing false alarms and detection lags. Outbreak detection can be implemented in different ways that range from simple predictors based on moving average filters  \cite{Farrington1996} and fusion methods \cite{Dubrawski2011} to complex spatial and temporal analyses \cite{Chan2011:cLUSTERING:SURVEILLANCE}, \cite{batlle2020adaptive}. 

In the early phases of a new pandemic, such as the recent Covid-19, the detection of recurring epidemic waves is particularly challenging because: (i) historical seasonal data are lacking, (ii) determining the current fraction of infected population can be difficult when many asymptomatic infected are present, and (iii) determining baselines and thresholds requires a precise characterization of the regional (time-varying) reproduction number.

\section{\NEW{Model--} Epidemiological models} \label{sec:models}

Mathematical epidemiology is a well-established field  aiming to model the spread of diseases both in human and animal populations \cite{rothman2008modern}, \cite{Martcheva2015}, \cite{thrusfield2018veterinary}. Given the high complexity of these phenomena, models are key to understand epidemiological patterns and support decision making processes \cite{Heesterbeek2015}.
There are in-host models that take into account the complexity of virus-host dynamics at the microscopic scale, describing how the pathogen interacts with cellular biomolecular processes and with the immune system, and between-host models that describe how the epidemic spreads within a population at the macroscopic scale, by considering the contagion either at an aggregate level (compartmental models) or through agent-based networked models of the population.
Approaches for epidemic multi-scale modelling, which include the interplay between immunological and epidemiological phenomena, are very recent and mostly rely on partial differential equations, sometimes reduced to small-size ordinary differential-equation systems, see e.g.
\cite{Almocera2018}, \cite{Almocera2019}, \cite{Barbarossa2015}, \cite{Cai2017}, \cite{Feng2015}, \cite{Gandolfi2015}, \cite{Gulbudak2020}, \cite{Hart2020}. Multi-scale epidemic modelling with an
interdisciplinary approach integrating epidemiology, immunology, economy and mathematics is advocated
in \cite{Bellomo2020}.

\subsection{Time-response and viral shedding}

In-host infection dynamics capture the interplay between virus and host. Models describing the dynamics of the immune response \cite{Castiglione2015} in the presence of an infectious disease have been proposed for influenza \cite{Handel2010}, \cite{Li2021}, \cite{Yan2017}, \cite{Zarnitsyna2016} and generic viral infections \cite{Moore2020}. Very recently an immunological description for Covid-19 has been provided \cite{Matricardi2020} and has enabled the characterization of virus-host dynamics for SARS-CoV-2 \cite{ABUIN2020}, \cite{HERNANDEZVARGAS2020}.

The evolution of a disease and its infectiousness over time can be characterized through some key epidemiological parameters (see e.g. \cite{heffernan2005perspectives},  \cite{wallinga2007generation}, \cite{vink2014serial},  and \cite{Hellewell2020}):

\begin{itemize}
    \item {\bf Latency time}: Time during which an individual is infected but not yet infectious. For Covid-19, initial estimates are of 3-4 days \cite{Li2020:virus}.
    \item {\bf Incubation time}: Time between infection and onset of symptoms. For Covid-19, the median incubation period is estimated to be 5.1 days, and 97.5\% of those who develop symptoms will do so within 11.5 days of infection \cite{Lauer2020:INCUBATION}; the median time from the onset of symptoms to death is close to 3 weeks \cite{Zhou2020:CLINICAL:COURSE}.
    \item {\bf Serial interval}: Time between symptom onsets of successive cases in a transmission chain \cite{vink2014serial}.
    For Covid-19, initial estimates of the median serial interval yield a value of around 4 days, which is shorter than its median incubation period \cite{Nishiura2020:Serial:Interval}; this implies that a substantial proportion of secondary transmission may occur prior to illness onset \cite{he2020temporal}.
    \item {\bf Infectiousness profile}: It characterizes the infectiousness of an infected individual over time. For Covid-19, the median duration of viral shedding estimation was 20 days in survivors, while the most prolonged observed duration of viral shedding in survivors was 37 days \cite{Zhou2020:CLINICAL:COURSE}.
    \item {\bf Basic reproduction number $\mathcal{R}_{0}$}: It represents the average number of new infections generated by an infectious person at the early stages of the outbreak, when everyone is susceptible, and no countermeasures have been taken \cite{heffernan2005perspectives}, \cite{wallinga2007generation}, \cite{liu2020reproductive}. For \NEW{the original strain of SARS-CoV-2}, first estimations range from 2.24 to 3.58 \cite{Zhao2020}; the effect of temperature and humidity in this parameter is addressed in different studies, see e.g. \cite{Mecenas2020:TEMPERATURE}. 
\end{itemize}

   The basic reproduction number, along with the serial interval, can be used to estimate the number of infections that are caused by a single case in a given time period. Without any control measure, at the early stages of the outbreak, more than 400 people can be infected by a single Covid-19 case in one month \cite{Nicola2020:REVIEW:Management}. Estimates of the basic reproductive number are of interest during an outbreak because they provide information about the level of intervention required to interrupt transmission and about the potential final size of the outbreak \cite{heffernan2005perspectives}.
     
The aforementioned parameters are often inferred from epidemiological models, once they have been fitted to available data on the number of confirmed infection cases and deaths \cite{rothman2008modern}, \cite{wallinga2007generation}. 

\subsection{Simple Compartmental models}\label{sec:basic:compartmental:models}
Compartmental models partition a population into different groups, called \textit{compartments}, associated with mutually exclusive stages of the disease. Each compartment is associated with a variable that counts the individuals who are in that stage of the infection \cite{Brauer2008}.

The simplest compartmental model\NEW{s are the SI, SIS, and SIR models}, introduced by Kermack and McKendrick at the beginning of the 20th century \cite{kermack1927contribution}. The SIR model includes three compartments: \textit{Susceptible} ($S$), representing healthy individuals susceptible of getting infected, \textit{Infected} ($I$), and \textit{Recovered}/\textit{Removed} ($R$). For possibly fatal diseases, this last compartment can take into account both recovered (with permanent immunity) and deceased individuals; however, for low mortality rate diseases, including only recovered individuals can be a good approximation. 

The SIR model describes the dynamics of an epidemic according to the following set of nonlinear differential equations:
\begin{eqnarray}
\frac{dS(t)}{dt} &=& -\beta S(t)I(t), \label{eq:SIR_S} \\
\frac{dI(t)}{dt} &=& \beta S(t)I(t) - \mu I(t),  \label{eq:SIR_I}\\
\frac{dR(t)}{dt} &=& \mu I(t),\label{eq:SIR_R}
\end{eqnarray}
where $\beta$ is the infection rate, while $\mu$ is the recovery rate; the variables $S$, $I$ and $R$ represent the fraction of susceptible, infected and recovered (or removed) individuals within the population, and $S(t)+I(t)+R(t)=1$ at all times $t$.
At the onset of a new epidemic, $S$ equals approximately the entire population, and thus from \eqref{eq:SIR_I} it holds that $I(t) = I_{0}e^{(\beta-\mu)t} =  I_{0}e^{\mu(\mathcal{R}_{0}-1)t}$, where $I_0$ represents the initial number of infected $I_0=I(0)$ and $\mathcal{R}_{0} = \beta / \mu$ is the \textit{basic reproduction number}, i.e. the average number of secondary cases produced by an infectious individual when $S\approx 1$. Clearly, when $\mathcal{R}_{0}$ is greater than 1, there is an exponential increase in the number of infected individuals during the early days of the epidemic. The same equation can also be used to estimate the point at which the number of newly infected individuals begins to decrease, $S(t)=1/\mathcal{R}_{0}$. At this point, the given population has reached what is known as \textit{herd immunity} \cite{fine2011herd}.

To account for the latency time, an extended version of the SIR model, called the SEIR model, includes an extra compartment for \textit{Exposed} (E) individuals, who have been infected but are not yet infectious, and are transitioning into the Infectious compartment at a fixed rate.


\subsection{Extended Compartmental models}\label{sec:Extended:compartmental:models}

To model the specific dynamics of a given infectious disease, extended compartmental models including additional compartments and transitions are often proposed. In particular, it is possible to consider symptomatic and asymptomatic compartments, vaccinated and unvaccinated, the possibility of reinfection after recovery, quarantined individuals, hospitalized, etc. Comprehensive books, surveys and works on compartmental models and their extensions are \cite{AndersonMay1991}, \cite{Brauer2012}, \cite{Breda2012}, \cite{Capasso1978}, \cite{Diekmann2000}, \cite{Gumel2004}, \cite{Hethcote2000}. 

\begin{figure}[ht!]
    \centering
    \includegraphics[width=0.45\textwidth]{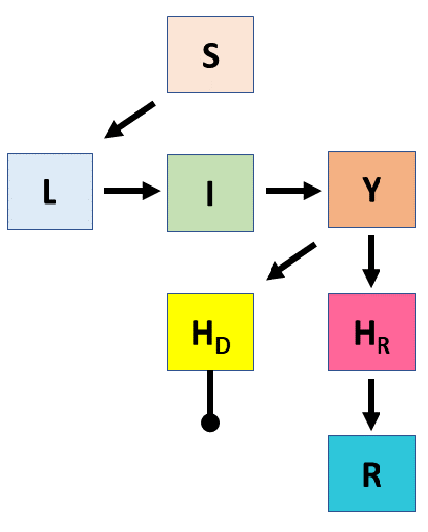}    \caption{Illustration of an extended compartmental epidemic model with seven compartments used in \cite{Riley2003} to model SARS : Susceptible (S), Latent (L), Asymptomatic and potientially infectious (I), Symptomatic Diagnosed (Y), Hospitalized that die ($H_D$), Hospitalized that recover ($H_R$) and Recovered R .}
    \label{fig:epidemic_model}
\end{figure}

The number of compartments required to model a disease depends on a variety of factors. For example, when modeling the dynamics of a new disease, for which no vaccine is available, it makes no sense to consider the vaccinated group. However, in other cases, as when modelling seasonal influenza, it is relevant to distinguish between vaccinated and unvaccinated populations \cite{Brauer2008}. Many diseases, like malaria, West Nile virus, etc., are transmitted not directly from
human to human but by infected animals (usually insects) \cite{taylor2001risk}. For these cases, the corresponding animal compartments are included in the model. Another relevant factor influencing what compartments to include in a model is the quantity and quality of available data. Complex models require more data to fit the parameters, so in the early \NEW{stages} of a new disease outbreak simple compartmental models are often employed.

Many applications of extended compartmental models can be found in the literature. For example, in \cite{Riley2003}, the authors use a dynamical compartmental model to analyze the effective transmission rate of the SARS epidemic in Hong Kong.  The model consists of 7 compartments: Susceptible individuals (S) become infected and enter a latent state (L). They then progress to a short asymptomatic and potentially infectious \NEW{state} (I) followed by a symptomatic state that leads to diagnose (Y) and hospitalization. It is assumed that every symptomatic case is eventually hospitalized and either dies ($H_D$) or, after treatment in the hospital ($H_R$), recovers (R) (see Figure \ref{fig:epidemic_model}). In \cite{Chowell2014}, a stochastic SEIR model is used to estimate the basic reproduction number of MERS-CoV in the Arabian Peninsula, distinguishing between cases transmitted by animals and secondary cases.

In the spread of the Covid-19 pandemic, asymptomatic infected individuals play a crucial role (see \cite{Ferretti2020} and \cite{Giordano2020}); the large prevalence of asymptomatic infections makes it harder to detect all cases and, thus, timely break the contagion chain. In \cite{Giordano2020}, a SIDARTHE model with eight compartments is proposed. This model distinguishes between asymptomatic and symptomatic infected, as well as detected and undetected infection cases, and partitions the population into Susceptible, Infected (asymptomatic infected, undetected), Diagnosed (asymptomatic infected, detected), Ailing (symptomatic infected, undetected), Recognised (symptomatic infected, detected), Threatened (infected with life-threatening symptoms, detected), Healed (recovered) and Extinct (dead) individuals. In \cite{Ferretti2020}, the epidemic model includes a transmission rate $\beta$ that takes into account the contributions of asymptomatic, presymptomatic and symptomatic transmissions, as well as environmental transmission. In both works, the results indicate that the contribution of asymptomatic infected to $\mathcal{R}_{0}$ is higher than that of symptomatic infected and other transmission modalities. In fact, symptomatic infected are often rapidly detected and isolated.

\subsubsection{Age-structured models} Age-structured epidemic models incorporate heterogeneous, age-dependent contact rates between individuals \cite{Valle2013:age:GROUPS}. In \cite{Xue-Zhi2001} and \cite{Safi2013}, stability results for different age-structured SEIR models are given. For Covid-19, an age-structured model, aiming at estimating the effect of physical distancing measures in Wuhan, is presented in  \cite{Prem2020}. In \cite{Saljeeabc3517}, a stratified approach is used to model the epidemic in France. 

\subsection{Seasonal behaviour}\label{sub:sub:sec:Seasonal}
Some works have studied the influence of temperature and humidity on the spread of viruses \cite{grassly2006seasonal}, \cite{waikhom2016sensitivity}. In the case of Covid-19, it has been reported that both variables have an effect on the basic reproduction number $\mathcal{R}_{0}$ \cite{Mecenas2020:TEMPERATURE}, \NEW{\cite{Effects:Climate:2020}}. This influence might be included in the epidemic models to capture the seasonal behaviour of Covid-19; for instance, by considering the parameters $\beta$ and $\mu$ as functions of both temperature and relative humidity. Yet, it remains unclear under which circumstances seasonal and geographic variations in climate can substantially alter the dynamics of a given pandemic, specially in the case of high susceptibility \cite{Baker2020:Seasonal:Limit}.

\subsection{Spatial epidemiology}
Compartmental models are well-suited to describe the evolution of epidemics in a single, well-mixed population where each individual is assumed to interact with every other at a common rate (homogeneous contacts). While this can be a reasonable approximation in some contexts, it is not appropriate to study the global spread of a pandemic over a large, geographically dispersed population. In the last decades, compartmental models have been successfully extended to spatial epidemiological models in order to analyze spreading phenomena where spatial patterns need to be more accurately described. Graphs and networks have often been used to achieve this, see for instance \cite{House2012}, \cite{Keeling2005}, \cite{Kiss2017}, \cite{Mei2017}, \cite{Nowzari2016:ieee:CONTROL}, \cite{PastorSatorras2015}, \cite{Lewien:Chapman:CDC:Focus}, \cite{ogura2016stability}, \cite{ogura2016epidemic},  \cite{Pare2020:ARC:BASAR}, \cite{Pare2018}, \cite{Zino2020}. \NEW{Three} widely used classes of models are described in the following sub-subsections.

\subsubsection{Meta-population models} Meta-population models integrate two types of dynamics: one related to the disease, typically driven by a compartmental model, and the other to the mobility of individuals (agent-based model) across the sub-populations that build the meta-population under analysis \cite{grenfell1997meta}, \cite{ball2015seven}. As a representative example, in \cite{Brockmann2013} the authors introduce the notion of effective distance to capture the spatio-temporal dynamics of epidemics, combining the SIR model of $n=1,2,\ldots ,p$ populations with mobility among them. The resulting model for each population is 
\begin{eqnarray*}\frac{dS_{n}(t)}{dt} &=& -\beta S_{n}(t)I_{n}(t) +  \sum_{m\neq n}{(w_{nm}S_{m} - w_{mn}S_{n})} \\
\frac{dI_{n}(t)}{dt} &=& \beta S_{n}(t) I_{n}(t) - \mu I_{n}(t) \\
&&+ \sum_{m\neq n}{(w_{nm}I_{m} - w_{mn}I_{n})},\\
\frac{dR_{n}(t)}{dt} &=& \mu I_{n}(t) + \sum_{m\neq n}{(w_{nm}R_{m} - w_{mn}R_{n})},
\end{eqnarray*}
where $w_{nm}$ is the per capita traffic flux from population $m$ to population $n$. In \cite{Aleta}, the authors use a SEIR compartmental model together with stochastic data-driven simulations to capture the mobility in all Spanish provinces. The work focuses on evaluating the effectiveness of \NEW{containment measures} in Spain on February 28th, when a few dozen cases of Covid-19 had been detected. Meta-population models to capture the spatio-temporal dynamics of the Covid-19 epidemics in Italy have been proposed in \cite{Bertuzzo2020}, \cite{gatto2020spread} and \cite{DellaRossa2020}. By capturing both temporal and spatial evolution of epidemics, meta-population models are also capable of forecasting the effectiveness of mobility restrictions.
\subsubsection{Social networks models} Social network models consider that transmission can only occur along linked or connected individuals \cite{El-Sayed2012}, which allows to explicitly model heterogeneity in contact patterns. Small-world networks have been used in combination with compartmental models to model disease transmission of SARS \cite{Small2005} and Covid-19 \cite{Thurner:social:network}, and also to assess the efficacy of contact tracing \cite{Kiss2006}. In general, network models produce a more accurate prediction of the disease spread \cite{Pare2020:ARC:BASAR}.  In particular, the use of homogeneous compartmental models in \NEW{populations} with heterogeneous contacts tends to underestimate disease burden early in the outbreak and overestimate it towards the end, although for certain kinds of networks compartmental models can be modified to prevent this problem \cite{Bansal2007}. Another interesting aspect of studying epidemic spreads with network models is the observation of a percolation phase transition \cite{House2012}, \cite{PastorSatorras2015}, i.e., an abrupt change in the global dynamics of the epidemics. Percolation theory has been widely studied in random networks
\cite{albert2002statistical}. In the context of epidemic modelling, the transition phase occurs where isolated clusters of infected people join to form a giant component that is able to infect many people \cite{harding2020phase}.     

\NEW{\subsubsection{Self-exciting spatio-temporal point processes  }}

\NEW{
In epidemiology, it is natural to register each new infection event with a pair $(t,x)$ in which $t$ refers to time and $x$ to location. The underlying stochastic model for this kind of data is called spatio-temporal point process \cite{diggle2006spatio:temporal:point:processes}. Since each infection event potentially causes new ones, an epidemic can be modelled as a self-exciting spatio-temporal point process in which the rate of infections depends on the past history of the process \cite{reinhart2018review}, \cite{Zino2020}. In this setting, the objective is to estimate an intensity function which predicts the rate of infections at any spatial location $x$ and time $t$  \cite{diggle2006spatio:temporal:point:processes}, \cite{LanceWaller2010:Chapterbook}. This modelling framework, which constitutes a generalization of Hawkes processes \cite{hawkes1971spectra}, permits the incorporation of the distributions of the duration of incubation, pre-symptomatic and asymptomatic phases, along with the modulating effect of time-varying counter-measures and detection efforts  \cite{GARETTO2021}.
}  \\

\subsection{Computer-based models}

Computer-based simulation methods to predict the spread of epidemics can take into account numerous factors, such as heterogeneous behavioural patterns, mobility patterns, both at long and short scales, demographics, epidemiological data, or disease-specific mechanisms \cite{marathe2013computational}, \cite{Helbing2015}. The real-world accuracy of mathematical
and computational models used in epidemiology has been considerably improved by the integration of large-scale data sets and explicit simulations of entire populations down to the scale of single individuals. These computational tools have recently gained importance in the field of infectious disease epidemiology, by providing rationales and quantitative analysis
to support decision-making and policy-making processes \cite{tizzoni2012real}.  As a representative example, the Global Epidemic and Mobility simulation framework (GLEAM) allows performing stochastic simulations of a global epidemic with different global-local mobility patterns, as well as data regarding demographics or hospitalization \cite{VanDenBroeck2011}.

However, detailed simulation-based methods depend on a significant number of parameters, which need to be chosen and fixed for a specific simulation. This is especially difficult in the early days of an epidemic outbreak. Furthermore, these approaches might not reveal which factors are actually relevant in the spread of epidemics. Simpler data-driven tools have also been developed to overcome these difficulties \cite{marathe2013computational}. 

\subsection{Modelling the effect of containment measures}\label{sub:modelling:measures}

Controlling an emerging infectious disease requires both the prompt implementation of countermeasures and the rapid assessment of their efficacy \cite{Cauchemez2006:Assessment:efficacy}, \cite{Gumel2004}, \cite{chowell2003sars}, \cite{Brauner2020:Effectiveness:Measures}, \cite{haug2020ranking}, \cite{GUAN:Prieur:FRANCE}. In what follows, we enumerate the most relevant non-pharmaceutical interventions, focusing on different research works that assess their efficacy.

\begin{itemize}

\item \textbf{Quarantine}: Quarantine of diagnosed cases, or probably infected, is crucial in every epidemic outbreak. In order to model the effect of quarantine, specific compartments are included in the epidemic models for SARs \cite{Gumel2004}, \cite{chowell2003sars}. If a significant fraction of the infected population is not diagnosed (or diagnosed with a significant delay), then the modelling is harder and non-diagnosed groups are included in the models \cite{may1987transmission}, \cite{Giordano2020}, \cite{Ansumali2020:ARC:Vidyasagar}.

Quarantine of a whole population (i.e., lockdown) is the most extreme measure in the scope of physical distancing/mobility restrictions. The extreme impact of Covid-19 yield to the quarantine of the epicentre of the pandemic (Wuhan) on January 24th, 2020, and the same measures were subsequently adopted in different countries of Europe and America \cite{gatto2020spread}. In this case, the effect of a lockdown can be modelled by means of time-varying epidemic models, see e.g. \cite{Calafiore2020time}.

\item \textbf{Physical distancing}: Physical (or social) distancing is another measure promoted by governments, public and private institutions in an attempt to reduce disease transmission \cite{Prem2020}, \cite{Morato2020:MPC:BRAZIL}, \cite{Prathereabc6197}. Population-wide wearing of masks, capacity reduction on public transport, reducing or stopping the activity in educational institutions or factories are examples of this. In \cite{Maharaj2012}, the authors conduct a simulation-based analysis to determine the effects of physical distancing both in public health and in the economy. Two social network models (regular and small-world networks) are combined with a compartmental SIR model, and the economic impact takes into account the costs of individuals falling ill and the cost of a reduction in social contacts. 

    \item \textbf{Mobility restrictions}:  Governments often introduce long-range or local mobility restrictions aimed at reducing disease transmission. Spatial epidemiology is particularly useful to model the effects of such measures. For instance, in \cite{Thurner:social:network}, the authors show, by means of a small-world network model, that the onset of mobility restrictions influences the final size of the outbreak, which is well below the levels of herd immunity.

\item \textbf{Proactive testing}: Proactive testing of asymptomatic individuals is very relevant for the monitoring and control of the Covid-19 pandemic \cite{WHO:Cases:and:Clusters}, since it allows to isolate infectious individuals and implement contact tracing strategies, which have been shown to be crucial \NEW{for} an effective control \NEW{of} the pandemic \cite{Giordano2020}.

\item \textbf{Contact tracing}: Contact tracing is a widely used epidemic control measure that aims to identify and isolate infected individuals by following the social contacts of individuals that are known to be infectious. A review of contact-tracing based epidemic models for SARS and MERS can be found in \cite{kwok2019epidemic}. In \cite{Kiss2006}, a small-world, free-scale network model is combined with a compartmental model to assess the efficacy of contact tracing.

\end{itemize}

\subsection{Fitting epidemic models to data}\label{subsection:Fitting:to:Data}

Dynamic epidemiological models rely on a set of parameters that have to be tuned in order to provide realistic predictions and/or infer essential features, such as the (time-varying) effective reproduction number \cite{Cori2013}, or the latent period. Fitting epidemic models to data is a fundamental problem in epidemiology that can be approached in different ways. We can distinguish between classical methods, in which the parameters of the model are unknown but fixed, and Bayesian methods, in which they are assumed to be random variables \cite{Kypraios2017}.  Another classification follows from the accessibility to the populations considered in the compartments of the model:
\begin{itemize}
    \item {\bf Full access} to the evolution of the number of cases in each compartment: In most models, the parameters that determine the dynamics multiply linear or bi-linear terms, depending on the current number of cases in each compartment. This means that a (vector) equality constraint, that depends (bi-)~linearly on the parameters to fit, can be obtained at each sample time. In the case of linear constraints, standard linear identification techniques, such as least-square methods, can be applied to estimate the parameters that best fit the model to the data. See, for example, \cite[Chapter 6]{Martcheva2015} and \cite{Allman2004}.
    \item {\bf Partial access} to the number of cases in each compartment: In many situations, there are no available time series for one or more of the groups considered in the model. This complicates the data-fitting process considerably because it is no longer possible to obtain, in a simple way, the equality constraints described in the full access case. The standard approach in this case is to resort to non-linear identification techniques (see \cite{Schon2011} and \cite{Abreu2020}). In this context, Monte Carlo based methods (e.g. Markov Chain Monte Carlo and Sequential Monte Carlo algorithms) play a crucial role in addressing the challenges that lie in reconciling predictions and observations \cite{McKinley2009}. 
\end{itemize}



\subsubsection{Sensitivity analysis}

Sensitivity analysis (SA) is the study of how the
uncertainty in the output of a model (numerical or
otherwise) can be apportioned to different sources
of uncertainty in the model input \cite{Saltelli2002:Sensitivity}.
See the review paper \cite{Qian2020:Sensitivity} on the use of this technique in the context of biological sciences. A monovariate and multivariate sensitivity analysis for a data-fitted SARS model is given in \cite{Alvarez2015:MORILLA}. The use of SA is common in many research papers on modelling Covid-19 (see e.g. \cite{Fang2020} and \cite{Saljeeabc3517}). 

\subsubsection{Validation and model selection}

The ultimate test of the validity of any model is that its behaviour is in accord with real data. Because of the simplifications introduced in any mathematical model of a biological system, we must expect
some divergence between the results of a model and reality, even for the most carefully collected data and most detailed model. Different questions arise in this context: (i) How can we determine if a model describes data well? (ii) How can we determine the parameter values in a model that are appropriate for describing real data? These questions are too broad to have a single answer \cite{Allman2004}, \cite{Vittinghoff2012}.

Epidemic models depend on their data calibration. However, many possible models are potentially suited to analyze the spread of a pandemic in a given moment. The models are inherently linked to the goal for which they were envisaged. For a given goal (for example second outbreak detection), different models can be considered. Model selection techniques are used on a regular basis in epidemiology \cite{Portet2020:AKAIKE:MODEL:SELECTION}. They address the problem of choosing, among a set of candidate models, the most suitable for a given purpose \cite{BurnhamKennethPandAnderson2010}. The selection is based on different aspects: (i) How the calibrated model is able to reconcile and match observations and (ii) the complexity of the model. Under similar adjustment to observations, simpler models are preferred since they are more robust from an information-theoretic point of view \cite{Huyvaert2011:REVIEW:MODEL:SELECTION}.
  
There are often different sets of parameters yielding a similar fit to data, but providing significantly different estimations of the main characteristics of the spread of the epidemic (like peak size, reproduction number, etc.). This issue is known as non-identifiability \cite{Roda2020:DIFFICULTIES}, \cite{Gustafsson:CDC:Focus:2020}. Identifiability issues may lead to inferences that are driven more by prior assumptions than by the data themselves \cite{Lintusaari2016:Identifiability}. There are some approaches to address this difficulty. The first one is to resort to simplified models (SIR and SEIR models, for example) in which the number of parameters to adjust is small, see \cite{Roda2020:DIFFICULTIES} and \cite{Postnikov2020:BACK:ENVELOPE}. The second one is to use data from different regions in a not aggregated way, which reduces the probability of parametric over-fitting \cite{fiacchini2020ockhams}. In this context, model selection theory provides systematic methodologies to determine which model structure best \NEW{suits} the purposes of the model \cite{BurnhamKennethPandAnderson2010}, \cite{Portet2020:AKAIKE:MODEL:SELECTION}.

\section{\NEW{Model--} Forecasting}\label{sec:forecasting}

The task of forecasting a time series can be stated as a supervised learning problem in which a number of temporal variables (also called predictors or \textit{features} in the machine learning literature) are used to learn a model able to predict the future value of an output variable of interest \cite{bishop2006pattern}. In our context, we focus on forecasting methods aiming to predict the future evolution of epidemiological variables \cite{suarez2017applications}, \cite{Chowell2019}. We find in the literature numerous approaches to forecast temporal variables describing the evolution of Covid-19 \cite{tayarani2020applications}, \cite{petropoulos2020forecasting}, \cite{Calafiore2020time}, from black-box approaches to estimates based on learning the internal parameters of compartmental epidemic models. Forecasting in the context of global pandemics faces many difficulties \cite{Ioannidis2020} and requires the implementation of validation and sensitivity analysis \cite{BurnhamKennethPandAnderson2010}.  We now introduce some considerations that should be taken into account in order to select and train a suitable forecasting model. 
 
 First, we start with some statistical considerations:
\begin{itemize}
    \item Frequentist versus Bayesian statistical methods: In the former, probabilities are assigned according to experiment repetition and occurrence. In the latter, the parameters of a model are learned using Bayes' theorem and prior knowledge about the probability distributions of unknown variables \cite{bonamente2013statistics}.
    \item Parametric versus non-parametric approaches: In the former, we assume a parametric function mapping past variables input into future predictions. This function contains several unknown parameters that are learned using historic time series. In the non-parametric approach, we do not assume such a parametric function \cite{malley2012probability}; for example, one can make future predictions for a given time series by analyzing the behavior of historic past behaviors resembling the behavior of the time series under consideration.
\end{itemize}

Other considerations to keep in mind are:

\begin{itemize}

    \item The model should be trained with reliable data.  If the available data is poor, the forecasts produced will be unreliable. In this direction, data-cleaning techniques such as data reconciliation, standardization, filtering, and outlier detection should be utilized to improve the quality of the input data collected \cite{Albuquerque1996}.
    \item The amount of data collected should be appropriate for the forecasting technique under consideration. For instance, black-box models, such as deep learning, require vast amounts of data compared with compartmental models \cite{Torrealba-Rodriguez2020}, \cite{Yang2020}; therefore, while dealing with relatively short time series, making predictions using compartmental models is more appropriate than using deep learning (and other black-box techniques).
    \item Learning procedures should include training, validation, and test phases executed separately. In other words, available data set should be divided into three parts, each one used for a different purpose. In the training stage, model parameters are learned using training data. In the validation step, one adjusts model hyper-parameters and performs comparisons with other competing approaches. Finally, the final test of a model should be carried out with data that has not been used during the training or validation phases \cite{BurnhamKennethPandAnderson2010}. 
    \item Interpretability of the model. While deep learning (and other black-box techniques) may produce high-quality predictions, the obtained model is hard to interpret; in other words, we typically do not have an intuitive understanding of why the model is making a prediction \cite{arrieta2020explainable}. However, when policy-makers make critical decisions based on the forecast of a model, it is important for them to understand why the model is behaving in a certain way. Therefore, it is sometimes reasonable to \NEW{use} more interpretable models, with parameters having a clear physical/biological interpretation, even at the expense of having a lower performance than \NEW{with} black-box approaches. 
\end{itemize}


\section{\NEW{Model--} Impact Assessment Tools}\label{sec:Impact:Tools}
In order to design effective control strategies, it is important to define the control goals first. 
In the context of the current pandemic, the ultimate goal is to maintain the spread of the virus within an adequate threshold (e.g., a low level of \NEW{infection cases} \cite{Priesemann2020}), while minimizing the economic and social impacts of the interventions. Once this goal is quantified in terms of a cost function, we should then consider the types of interventions that can be taken to achieve our goals, as well as their associated costs. For example, there are several non-pharmaceutical interventions that can be used before a vaccine is widely available, such as physical distancing, border closures, school closures, isolation of symptomatic individuals, among others (see Section \ref{sub:modelling:measures}). \NEW{Each of these interventions has} an associated economic and social cost that should be considered while making a decision.

In order to use disciplined decision-making techniques, like the ones described below, one needs to clearly state the control objectives in a precise, quantitative form. Furthermore, it is necessary to quantify the impact and costs of all possible interventions, as well as their actuation limits \cite{Cauchemez2006:Assessment:efficacy}, \cite{Brauner2020:Effectiveness:Measures}. In this direction, we can quantify the impact of our actions by using suitable indexes such as the mean reproductive number, the mortality index, or the unemployment rate or public debt, to name just a few. Once the decision-maker has decided how to use these indexes to measure the impact and cost of potential actions, the decision-making process can be stated as a formal optimization problem with constraints. For example, the goal could \NEW{be} the minimization of a weighted index measuring the economic and social impact of our actions while keeping the reproductive number smaller than one.

We would like to remark that the numerical estimation of certain indexes is not an easy task because they require the design of data-driven strategies to assess the effect of each potential decision on different indexes. This could be done by means of predictive models and forecasting schemes analyzed in the previous sections.  In some cases, quantifying the effect of one intervention over the spread of an epidemic is a non-trivial task, since multiple interventions are typically present at the same time \cite{OHANNESHAUSHOFER2020}. In these scenarios, correlation analyses, like Pearson Correlation Coefficient (PCC), can be a naive way to assess causalities. \NEW{Whenever} possible, a reliable approach to establish causalities is to perform Randomized Control Trials (RTC) \cite{DONNER:Randomized:Trials:94}, \cite{OHANNESHAUSHOFER2020}. In an RTC, a subset of randomly chosen individuals receives an intervention, while the rest of individuals receives no intervention. A standard statistical analysis of the observed results can be used to \NEW{reliably} evaluate the impact of this intervention. In the following subsections, we discuss a collection of indexes that could be included in the decision-making process of managing a pandemic. 

\subsection{Spread of the virus and reproductive number}
It is natural to express the effectiveness of control strategies in terms of the effective reproductive number $\mathcal{R}(t)$. As introduced in Section \ref{sec:models}, the basic reproduction number $\mathcal{R}_{0}$ determines the potential of an epidemic to spread exponentially at its early stage by measuring the number of secondary infections induced by a typical infectious individual in a population when everyone is susceptible. In contrast, when an epidemic is \NEW{ongoing}, the effective reproduction number, denoted by $\mathcal{R}(t)$, is used to quantify the average number of secondary \NEW{infections} per infectious case in a population \NEW{with} both susceptible and non-susceptible hosts. The effective reproduction number can be used to assess the ability of available control measures to contain the spread of an epidemic. By implementing interventions able to maintain $\mathcal{R}(t)$ below 1, the incidence of new infections decreases and the spread of epidemics fades with time. In \cite{Cori2013}, the authors presented a software tool that was validated with 5 different epidemics, including SARS and influenza. This tool can be used to estimate the daily reproductive number \NEW{$\mathcal{R}(t)$} and its variation in the presence of vaccination and super-spreading events. 

For Covid-19, a numerical analysis of the effective reproductive \NEW{number} can be found in \cite{Fang2020}, where, using real data and a SEIR model, the authors estimate $\mathcal{R}(t)$ in Wuhan and quantify the effectiveness of government measures. Based on the number of deaths, in \cite{Flaxman2020}, the Imperial College Covid-19 Response Team used a semi-mechanistic Bayesian model to estimate the evolution of $\mathcal{R}(t)$ when non-pharmaceutical measures, such as physical distancing, self-isolation, school closure, public events banned, and complete lock-down, were recommended/enforced.

Limitations in the use of $\mathcal{R}(t)$ as an assessment tool stem from the unreliability of available data sources. As a result, determining the real value of $\mathcal{R}(t)$ is difficult. Other indirect measures, like the number of deaths, ICU cases, saturation of healthcare systems can also be employed to assess the current epidemic burden, as described in the next subsection.

\subsection{Healthcare systems capacity}
The capacity of a country to prevent, detect, and respond to epidemic outbreaks \NEW{varies} widely across countries. The preparedness and resilience of a healthcare system is a particularly relevant factor to analyze the future impact of an infectious outbreak in the population \cite{kandel2020health}. The capacity of a healthcare system to continue delivering the same level (quantity, quality and equity) of basic healthcare services and protection to the population can severely degrade during an epidemic outbreak \cite{blanchet2017governance}, \cite{Emanuel2020}.  At the early stages of the Covid-19 outbreak, its virulence and high contagiousness quickly saturated the healthcare system of many cities around the world, resulting in higher mortality rates \cite{Miller2020}, \cite{Lai2020}. Furthermore, in countries with low capacity, like African and South American countries, saturation levels are reached even with a significantly smaller number of cases \cite{velavan2020covid}, \cite{Morato2020:MPC:BRAZIL}. 

To limit the saturation of healthcare systems and plan resource distribution effectively, tools that assess the effect of different interventions on the magnitude and timing of the epidemic peak during first and secondary outbreaks (see Sections \ref{sec:models} and \ref{sec:forecasting}) are fundamental. However, precise tools to forecast these peaks are challenging to obtain, due to the limitations of the available data and the time-varying nature of the mitigation efforts and potential seasonal behaviour of a pandemic. Another issue is the uncertain adherence of the population to the interventions (see next subsection). In order to partially circumvent these issues, forecasts of cumulative disease burden are often looked for. While missing the intensity and timing of the peaks, these projections can at least allow to identify areas with heavy present and/or future pandemic incidence.

\subsection{Adherence to interventions and social impact }

Analyses of the relationship between risk perception and preventive behaviours can be found  in the social epidemiology literature \cite{berkman2014social}, \cite{Zan:Emotional:Responses:Covid19}. Moreover, the level of belief in the effectiveness of recommended behaviours and trust in authorities are important predictors of adherence to preventive behaviour  (see \NEW{the} survey paper \cite{bish2010demographic:behaviour}), which is fundamental to deploy effective containment strategies \cite{moran2016epidemic}. Here, we review some of the methodologies that could be helpful to design indexes aiming to monitor the adherence of the population to interventions and the social burden of the pandemic.

\begin{itemize}
    \item \textbf{Social network analysis}: Online social networks, such as Facebook and Twitter, can be used to assess the impact of an infectious disease on society. People post in these social networks their feelings and worries. In \cite{shanthakumar2020understanding}, 530,206 tweets in the USA were analyzed to measure the social impact of Covid-19. The hashtags were classified into six categories, including general covid, quarantine, school closures, panic buying, lockdowns, frustration and hope. Thus, the number of tweets in each category can be used as a metric of social impact and overall sentiment. Similarly, Weibo microblogging social network was used in \cite{li2020characterizing} to study the propagation of situational information related to Covid-19 in China. In \cite{jiang2020political}, the political polarization with regards to Covid-19 in the United States was analyzed using a large Twitter dataset.
    \item \textbf{Search engines}: Online searches made by citizens in search engines, such as Google, Bing, or Baidu, can be used to measure the social impact of the epidemic in different locations. Normally, people try to find information \NEW{about} unknown diseases, drugs, vaccines, and treatments on the Internet. Along this line, the authors of \cite{Ginsberg2009} found a correlation between the relative frequency of certain queries in Google and the percentage of physician visits in which a patient presents influenza-like symptoms. Furthermore, other works have performed similar studies for other epidemics like Influenza Virus A (H1N1) \cite{Cook2011}. Regarding Covid-19, in \cite{Qin2020}, the Baidu engine is used to estimate the number of new cases of Covid-19 in China by the number of searches of five keywords, such as dry cough, fever, chest distress, coronavirus, and pneumonia. These five keywords showed a high correlation with the number of new cases. 
    \item \textbf{News}: The number and the content of posts in online newspapers can also be used to assess the spread of the virus. Along this line, in \cite{zheng2020predicting}, Natural Language Processing (NLP) is used to extract the relevant features of news media in China.
    \item \textbf{Online questionnaires}: Another tool for measuring the social impact of a sanitary emergence is through online questionnaires such as \cite{oliver2020covid19impact} (Spain,  146,728 participants), \cite{qiu2020nationwide} (China, 52,730 participants) and \cite{kleinberg2020measuring} (UK, 2,500 participants), which were implemented for the Covid-19 pandemic. These questionnaires allow to rapidly ask citizens multiple questions related to adherence to interventions, as well as psychological, social and economic impact,  among other aspects. The main difficulty is to spread the questionnaires throughout the population, although social networks and web-based tools help to reach a large amount of population.
    \item \textbf{Mobility}: 
    One of the most relevant indexes to understand the spread of a pandemic is mobility \cite{tizzoni2014use}. See Section 8.4 in \cite{Alamo2020b} for a relation of mobility data sets in the context of Covid-19. The
    reduction of mobility is not only due to the imposed quarantines and lockdowns by governments but also \NEW{due to} the increasing population's fear of getting infected. In \cite{Engle2020}, a perceived risk index of contracting Covid-19 is defined. This metric measures the individuals’ perception of risk, and it is determined by several variables, such as prevalence in both local and neighbouring locations, as well as population demographics. The results in \cite{Engle2020} indicate that a rise of local infection rate from 0\% to 0.003\% reduces mobility by 2.31\%.
\end{itemize}

\section{\NEW{Manage--} Managing and Decision Making}\label{sec:Decision:Making}


Deciding which of the far-reaching social and economic restrictions are \NEW{the} most effective to contain the spread of a disease, as well as the conditions under which they can be safely lifted, is one of the main goals of data-driven decision approaches to combat pandemics. Unlike an unmitigated pandemic, which spreads through the susceptible population out of control and eventually fades out, a mitigated pandemic presents waves. For example, a first wave grows when a very transmissible disease appears and decreases due to, for example, social distancing measures. However, as soon as social distancing measures are relaxed, a new wave can appear as long as we have a large number of individuals susceptible to the infection. To avoid \NEW{recurrent} waves, it is important to put in place surveillance systems and reactive mechanisms to reduce the potential burden of secondary epidemic waves. The decision-making process in this context is complex for many reasons:
\begin{itemize}
    \item The presence of uncertainty in some crucial parameters characterizing the spread, such as seasonality, extent and duration of immunity of a new pandemic outbreak \cite{Cobey2020:SEASONAL}, \cite{Kissler2020:POST:PANDEMIC}.   
    \item The difficulties in assessing the quantitative effect of a specific set of mitigation interventions on the effective reproduction number \cite{OHANNESHAUSHOFER2020}. 
    \item The possibility of significant non-symptomatic transmission (as in the case of Covid-19), which renders some interventions less effective \cite{Nishiura2020:Serial:Interval}, \cite{Prathereabc6197}.
    \item The different regional incidence and adherence to interventions, which motivates spatially distributed decisions \cite{Selley2015}, \cite{DellaRossa2020}.
    \item The limited capacity of healthcare systems and the logistic challenges to address mass testing and mass vaccination.
    \item The necessity to mitigate the spread of the epidemic and, at the same time, \NEW{reduce} the socioeconomic impact. 
    \item The time-delay induced by the incubation period of the disease, as well as the testing system, which does not allow for a prompt evaluation of the effect of the implemented actions. 
    \item The difficulties of assessing in a quantitative way the disruptive effects of the undertaken measures on relevant macroeconomic variables.
    
\end{itemize}

In what follows, we analyze under which circumstances the epidemic can be mitigated (controllability of the pandemic). After that, we also discuss some methodologies that have been applied to combat infectious diseases, including the Covid-19 pandemic, and that could potentially be applied in the context of future pandemics. See also the review papers \cite{Nowzari2016:ieee:CONTROL}, \cite{bussell2019applying} for the use of control theory in the context of disease control, or \cite{preciado2014optimal}, \cite{Pare2020:ARC:BASAR}, \cite{Ansumali2020:ARC:Vidyasagar} for the stability analysis of an epidemic. 

\subsection{Controllability of the pandemic}\label{sec:Controllability}

In this subsection, we review the most important factors determining the controllability of a pandemic: the aspects that have a relevant impact on the effective reproduction number. We link them with standard epidemic threshold theorems (e.g. \cite{Becker1977}, \cite{Whittle1955}, \cite{kermack1927contribution}). 

The epidemic threshold theorem of Kermack and McKendrick \cite{kermack1927contribution}, stated in 1927, and in particular its stochastic form as given by Whittle \cite{Whittle1955} are fundamental to predict the size and nature of an infectious disease outbreak. The
theorem indicates that, in homogeneously
mixed communities, major epidemics can be prevented by keeping the product of the size of the susceptible
population, the infection rate, and the mean duration of the infectious
period, sufficiently small \cite{Becker1977}. 
We now discuss how to have an impact on each of these factors by means of control actions.

\begin{itemize}
\item {\bf Size of the susceptible population:} The most effective way to reduce the susceptible population is by means of vaccines: vaccination campaigns increase herd immunity to a level that prevents further spread of the disease \cite{Sherer:Mass:Vaccination}, \cite{Giordano2021}. Protection against an infectious disease can either be achieved by widespread vaccination or by repeated waves of infection over the years, until a large enough fraction of the population is immunized \cite{Graham2020:VACCINE}. However, an issue is the duration of the acquired immunity \cite{Kissler2020:POST:PANDEMIC}, which in some infectious diseases, like the seasonal influenza, is not long enough to prevent recurring seasonal peaks \cite{Cobey2020:SEASONAL}.   
\item {\bf Infection rate:} This factor can be reduced by means of different control actions like physical distancing, mobility constraints or prohibition of certain activities \cite{Kraemer2020}, \cite{Ngonghala2020:control:assessment}. Depending on the seasonality and the specific demographic characteristics of a given population, the implemented measures can exhibit a time-varying effect on the infection rate \cite{Cori2013}, \cite{Fang2020}. This might cause flows from tropical to temperate regions and back in each hemisphere’s respective winter,  limiting opportunities for global \NEW{disease} declines \cite{Cobey2020:SEASONAL} and implying that surveillance methods to detect a seasonal peak should be put in place.
 
\item {\bf Mean  duration of the infectious period:} 
An effective way to reduce the infectious period consists in detecting infected cases and setting them into quarantine \cite{chowell2003sars}. 
Challenges are posed by relatively short latent periods and by the presence of many asymptomatic cases, as in the Covid-19 pandemic; then, the impact of quarantine measures depends very much on how fast the detection is taking place. It has been shown that the \NEW{probability of effectively controlling the outbreak} decreases with long delays from symptom onset to isolation \cite{Hellewell2020}, \cite{Ferretti2020}. A large prevalence of asymptomatic cases is indeed an issue due to the significant probability that transmission occurs before the onset of symptoms (\NEW{when the} median latent delay is smaller than \NEW{the} median incubation time), hence before the infection can be detected \cite{Ferretti2020}, \cite{Giordano2020}. 
\end{itemize}



\subsection{Optimal allocation of limited resources}\label{Optimal:resource:Allocation}

During a major health crisis, policy makers face the problem of optimally allocating limited resources, such as intensive care beds, ventilators, tests, high-filtration masks and Individual Protection Equipment (IPE), medicines, vaccines, etc. \cite{Brandeau2003}, \cite{Zaric2002}. 
This fact has led to the problem of how to ethically and consistently allocate resources \cite{Emanuel2020}. In this context, the term ``resource allocation problem'' extends to issues such as where and when to allocate available resources.

A rigorous and precise allocation method should lead to the formulation of an optimization problem, composed of a mathematical formulation and efficient algorithms to obtain its numerical solution \cite{Hansen2011}. In the mathematical model, resource allocations are the decision variables while the objectives are encoded in cost functions and equality or inequality constraints. For example, in \cite{Zaric2002} and \cite{Brandeau2003}, budget allocation models for multiple populations are provided. In \cite{preciado2013optimal}, a network model is used to optimally allocate vaccines to eradicate an initial epidemic outbreak using linear matrix inequalities. An extension of this work to the case of directed and weighted networks can be found in \cite{preciado2014optimal} and \cite{nowzari2015optimal}, where geometric programming was proposed to find an optimal solution. The same authors extend this last result to more general compartmental models in \cite{Nowzari2017}. See also \cite{hayhoe2020data} for an application of geometric programming and multi-task learning in the context of Covid-19.

In \cite{Lampariello2020} an optimization problem is formulated to find the number of tests to be performed in the different Italian regions in order to maximize the overall detection capabilities. The problem is a quadratic, convex optimization program. In \cite{Gollier2020}, a group testing \cite{Walter:1980:Poolin} approach is considered, and it is shown how the optimization of the group size can save between 85\% and 95\% of tests with respect to individual testing. See also \cite{Yilmaz2020:FOCUS} for \NEW{a} strategy that optimizes testing resources in the context of the Covid-19 pandemic.

Estimation, forecasting, and impact assessment techniques are often used to allocate resources, as they enable decision-makers to predict imbalances between supply and demand and to evaluate the overall efficiency of different alternatives of allocation. In \cite{Emanuel2020}, the authors propose fair resource allocation guidelines in the time of Covid-19, which can be a reference for future pandemics. These guidelines come from four fundamental values: (i) maximizing the benefits, (ii) treating people equally, (iii) promoting instrumental value, and (iv) giving priority to the worst off. As a result, these guidelines are condensed in some recommendations:

\begin{enumerate}
    \item To maximize the number of saved lives and life-years, with the latter metric subordinated to the former.
    \item To prioritize critical interventions for healthcare workers and others who take care of sick patients because of their instrumental value.
    \item For patients with similar prognoses, equality should be invoked and operationalized through random allocation.
    \item To distinguish priorities depending on the interventions and the scientific evidence (e.g. vaccines could be prioritized for older persons while allocation ICU resources depending on prognosis might mean giving priority to younger patients).
    \item People who participate in research to prove the safety and effectiveness of vaccines and therapeutics should receive some priority for interventions.
\end{enumerate}

\subsection{Trigger Control}\label{sec:Trigger:control}

A strategy to modulate the intensity of non-pharmaceutical interventions consists in implementing a trigger mechanism to maintain the effective reproduction number close to one, avoiding the saturation of the healthcare system while reducing, when possible, the economic and social burden of the pandemic \cite{Cauchemez2006:Assessment:efficacy}, \cite{preciado2014optimal}, \cite{DellaRossa2020}, \cite{bin2020fast}.  
The on-line surveillance of the pandemic permits to estimate the time-varying value of the effective reproduction number. Three cases are possible:

{\it The effective reproduction number is largely under 1:} in this case, one could consider lifting one, or more non-pharmaceutical measures. However, other criteria should be met in order to implement a reduction on the confinements measures in a safe way \cite{Priesemann2020}. The three criteria highlighted by the European Commission to decide on the lifting of confinement measures for Covid-19 \cite{EuropeanCommission2020:Lifting} are: 
\begin{enumerate}
    \item Epidemiological criteria showing that the spread of the disease has significantly decreased and stabilised for a sustained period of time. This can, for example, be indicated by a sustained reduction in the number of new infections, hospitalisations and patients in intensive care.
    \item Sufficient health system capacity, in terms of, for instance, occupancy rate for Intensive Care Units; adequate number of hospital beds; access to pharmaceutical products required in intensive care units; reconstitution of stocks of equipment; access to care, in particular for vulnerable groups;  availability of primary care structures, as well as sufficient staff with appropriate skills to care for patients discharged from hospitals or maintained at home and to engage in measures to lift confinement (testing for example). This criterion is essential as it indicates that the different national healthcare systems can cope with future increases in cases after lifting the measures. At the same time, hospitals are likely to face a backlog of elective interventions that had been temporarily postponed during the pandemic peak. Therefore, healthcare systems should have recovered sufficient capacity in general, and not only related to the management of Covid-19.
    \item Appropriate monitoring capacity, including large-scale testing capacity to detect and monitor the spread of the virus combined with contact tracing and possibilities to isolate people in case of \NEW{resurgence} and further spread of infections. Antibody detection capacities, \NEW{e.g. in the case of Covid-19, provide} complementary data on the share of the population that has successfully overcome the disease and eventually measure the acquired immunity.
\end{enumerate}
 {\it The effective reproduction number has increased to a level clearly above 1:} this would demand, in most cases, extremely prompt strengthening of the mitigation interventions. The stringency of the new measures should guarantee that the healthcare system is not overwhelmed by a new epidemic wave. This requires the implementation of forecasting tools that help decision-makers to determine the most suitable set of mitigating measures. 
 
 {\it The effective reproductive factor is close to 1:} in this case, a deeper analysis is required. The decision on whether to keep the same set of current mitigation measures or not will depend on the current fraction of infected population, the healthcare system capacity, and the potentiality of implementing in a short period of time a mitigating intervention, which is capable of bringing the effective reproductive number to admissible values. That is, the decision could be determined by the worst-case cost of delaying in one week the implementation of new measures. It is worth stressing that preemptive actions are always preferable: the earlier a countermeasure is adopted, the better in terms of its efficacy and potential to save lives \cite{Giordano2021}.


In order to develop a timely and appropriate response, different methodologies from the field of control theory are available (see the review paper  \cite{Nowzari2016:ieee:CONTROL}).
Relying on Pontryagin’s maximum principle, optimal control approaches have been proposed to design optimal treatment plans, or vaccination plans, that minimize the cost of the epidemics, including both the cost of infection and the cost of treatment or vaccination \cite{Bloem2009} \cite{Forster2007} \cite{Hansen2011} \cite{Morton1974}.
Robust control approaches have also been proposed to control the spreading of infectious diseases, seen as uncertain dynamical systems \cite{LeeLeitmann1994} \cite{Leitmann1998}.
We provide more details on optimal control approaches in the following subsections.

\subsection{Optimal Control Theory}\label{sec:Optimal:Control:Theory}

Optimal control theory \cite{Liberzon2012:optimal:control} can be applied to reduce in an effective way the burden of an epidemic \cite{Lenhart2007:OPTIMAL:CONTROL}, \cite[Chapter 9]{Martcheva2015}. The dynamic optimization techniques of the calculus of variations and of optimal control theory provide methods for solving planning problems in continuous time. The solution is a continuous function (or a set of functions) indicating the optimal path to be followed by the variables through time or space \cite{KamienMortonIandSchwartz2012:optimal:control}. 
We present here a common formulation of a continuous dynamical optimization problem \cite[Section 2]{Hartl1995:OPTIMAL:CONTROL}:
\begin{eqnarray}
\min\limits_{x(\cdot),u(\cdot)} &&S(x(T),T)+ \int\limits_0^T F(x(T),u(t),t)dt \nonumber\\
s.t. && x(0)=x_0, \nonumber \\  
&& \dot{x}  =  f(x(t),u(t),t), \nonumber\\
&& g(x(t),u(t),t) \geq 0, \label{ineq:mixed}\\
&& h(x(t),t) \geq 0, \label{ineq:pure}\\
&& a(x(T),T) \geq 0, \label{ineq:Terminal}\\
&& b(x(T),T)  =  0. \label{equ:Terminal}
\end{eqnarray}
In an epidemic control problem $x(t)$ represents the state of the pandemic at time $t$ (for example, in terms of the populations of the different compartments), $u(t)$ is the control action which can be stated in a direct way (intensity of the interventions, number of vaccines, treatments), or in an indirect way (infection rate, immunologic protection, recovery rate). The differential equation $\dot{x}(\cdot)=f(\cdot,\cdot,\cdot)$ represents the epidemic model, inequality (\ref{ineq:mixed}) allows us to incorporate (mixed) constraints on $x(\cdot)$ and $u(\cdot)$ whereas the (pure) constraint (\ref{ineq:pure}) can be used to impose limits on the size of the components of $x(\cdot)$. Finally, (\ref{ineq:Terminal}) and (\ref{equ:Terminal}) are terminal constraints. The question of existence of optimal pairs $(x^*(\cdot),u^*(\cdot))$ was studied in \cite{Cesari1965:Existence:OPTIMAL:CONTROL} and \cite{Filippov1962:EXISTENCE:OPTIMAL:CONTROL}. See also \cite[Section 3]{Hartl1995:OPTIMAL:CONTROL} and the references therein.  

Pontryagin's maximum principle provides necessary conditions that characterize the optimal solutions in the presence of inequality constraints \cite{Liberzon2012:optimal:control}, \cite{Kirk2004:OPTIMAL:CONTROL}. These necessary conditions become sufficient under certain convexity conditions on the objective and constraint functions \cite{Mangasarian1966:OPTIMAL:CONTROL}, \cite{Kamien1971:OPTIMAL:CONTROL}. In general, the solution of the optimal problem in the presence of nonlinear dynamics and constraints requires iterative numerical methods to solve the so-called Hamiltonian system, which is a two-point boundary value problem, plus a maximum (minimum) condition of the Hamiltonian (see e.g. \cite[Chapter 6]{Kirk2004:OPTIMAL:CONTROL}).  

We now describe some examples of the use of optimal control theory in epidemic control. In \cite{Zaman2008:OPTIMAL:CONTROL},
the dynamic optimal vaccination strategy for a SIR epidemic model is described. The optimal solution is obtained using a forward-backward iterative method with a Runge-Kutta fourth-order solver. An example of how to deploy scarce resources for disease control when epidemics occur in different but interconnected regions is presented in \cite{Rowthorn2009:optimal:control}. The authors solve the optimal control problem of minimizing the total level of infection when the control actions are bounded. 

In \cite{Youssef2013} the authors apply Pontryagin's Theorem to obtain an optimal Bang-Bang strategy to minimize the total number of infection cases during the spread of SIR epidemics in contact networks. Optimal control theory is employed to design the best policies to control the spread of seasonal and novel A-H1N1 strains in \cite{Prosper2011:OPTIMAL:CONTROL}. An example of the use of optimal control theory to control the present Covid-19 pandemic is presented in \cite{Mandal2020} and \cite{hayhoe2020data}, where the authors design an optimal strategy, for a five compartmental model, in order to minimize the number of infected cases while minimizing the cost of non-pharmaceutical interventions.

\subsection{Model Predictive Control}\label{sec:MPC}

Model predictive control (MPC) provides optimal solutions to a control decision problem subject to constraints \cite{Camacho2013:MPC}, \cite{rawlings2017:MPC}.  MPC is a receding horizon methodology that involves repeatedly solving a constrained optimization problem, using predictions of future costs, disturbances, and constraints over a moving time horizon. In epidemic control, the aforementioned optimization problem is solved daily, or weekly, in order to decide the optimal control action (for example, the intensity of mitigation interventions, or the optimal allocation of resources). The output of the model  predictive controller is adaptive in the sense that it takes into consideration the latest available information on the state of the pandemic \cite{selley2015dynamic}, \cite{bussell2019applying}. See, for example, \cite{Morato2020:MPC:BRAZIL}, \cite{Kohler2020:mpc:with:FRANK}, \cite{Alleman2020:MPC}  for MPC formulations that address the control of the Covid-19 pandemic. See \cite{carli2020model} for a review paper on the application of MPC in the context of Covid-19 pandemic.

Because of the spatial clustered distribution of an epidemic, it is possible to apply specific control techniques from the field of distributed model predictive control \cite{Maestre2014}, \cite{Christofides2013}. For example, non-linear model predictive control can be used to control the epidemics by solely acting upon the individuals’ contact pattern or network \cite{Selley2015}.
Another example of distributed MPC in the control of epidemics is given in  \cite{Kohler2018:MPC:Resource:Allocation},  where the problem of dynamically allocating limited resources (vaccines and antidotes) to control an epidemic spreading process over a network is addressed.  

\subsection{Multi-objective control}

Pareto optimality is used in multi-objective control problems with counter-balanced objectives. For instance, in a counter-balanced bi-objective problem, improving one objective implies to worsen the other one. Pareto optimality is based on the Pareto dominance, which defines that one solution dominates another one if it is strictly superior in all the objectives. Thus, the goal of the optimization algorithm is to find the Pareto front, which includes all dominant solutions of the control problem. Therefore, there is a set of optimal solutions instead of one optimal solution. The Pareto front is a useful tool for decision-makers that allows to visualize all the possible optimal solutions (for two objectives is a curve, for three objectives a plane, and so forth) and to evaluate the trade-off between different strategies. In the context of epidemic control \cite{Sharomi2017}, Pareto optimality has been used in \cite{Yousefpour2020} in a bi-objective control problem, the goals are related to epidemic measures like the number of cases and economic costs.     


\section{Conclusions}\label{sec:Conclusions}

This \NEW{review} has presented a roadmap for controlling present and future pandemics from a data-driven perspective, based on three pillars: Monitoring, Modelling, and Managing. We have highlighted the interplay between data science, epidemiology, and control theory to address the different challenges raised by a pandemic. 

Methodologies and approaches proposed for previous epidemics and the present Covid-19 pandemic have been reviewed, without claiming exhaustiveness, given the huge and continuously growing literature on this subject. Although the relevant body of literature is extremely large and many approaches have been studied in the past, further research is still needed. Implementing effective control strategies to mitigate a pandemic is difficult because of various reasons:  (i) the unavoidable uncertainty \NEW{affecting} some crucial parameters that characterize the spread, including compliance issues due to the unpredictable human behaviour, (ii) the difficulties in assessing the quantitative effect of mitigating interventions, (iii) the impossibility of obtaining a prompt evaluation of the effect of the implemented interventions, due to the intrinsic time-delay, and at the same time the critical importance of acting quickly, due to the exponential nature of the spreading phenomenon: even a small delay in interventions can lead to a much heavier healthcare burden and a much larger death toll \NEW{(see e.g. \cite{Giordano2021})}.

The first step for modelling different aspects of the pandemic is the processing of the available raw data to obtain consolidated time-series.  In order to obtain predictive models, which are crucial for the decision-making process, we have discussed several techniques from epidemiology and machine learning. We have described the most relevant modelling and forecasting approaches, focusing on the adjustment of the prediction models to the available data, model selection and validation processes.

Different surveillance systems able to detect, or anticipate, possible recurring epidemic waves have been surveyed. These systems enable an immediate response that reduces the potential burden of the outbreak.  Different methods from control theory can be applied to provide an optimal, robust and adaptive response to the time-varying incidence of an epidemic. These methods can be applied to the optimal allocation of resources, useful for testing campaigns and vaccination plans, and to determine trigger control schemes that modulate the stringency of the adopted interventions.  
We have reviewed the control-theory literature focused on the analysis and the design of feedback structures for the efficient control of an epidemic. Besides, we have also mentioned some techniques from distributed model predictive control that can be applied to control the temporal and spatial evolution of an infectious disease.

Preventing and controlling pandemics will be an increasingly important challenge in the future, due to the likelihood of new virus spillovers resulting from the increasing ecological footprint of humans. The systems and control community has powerful tools available to contribute and take on this fundamental challenge.


\bibliography{Bib_Survey_Data_Driven}

\end{document}